\documentclass{ws-procs9x6}

\begin{document}

\title{Disordered complex systems using cold gases and trapped 
ions\footnote{\uppercase{T}his work is partially supported by \uppercase{D}eutsches \uppercase{F}orchunggemainshaft \uppercase{SFB}407, 
\uppercase{SPP}1078 and \uppercase{SPP}1116, the  \uppercase{S}panish \uppercase{MCYT BFM}-2002-02588 and \uppercase{MAT}2002-00699, the 
\uppercase{A}lexander von \uppercase{H}umboldt \uppercase{F}oundation, the \uppercase{EC P}rogram
\uppercase{QUPRODIS}, the \uppercase{ESF P}rogram \uppercase{QUDEDIS}, and \uppercase{EU IP SCALA}.
}}

\author{Aditi SEN(DE), Ujjwal SEN, AND Maciej LEWENSTEIN\footnote{\uppercase{A}lso at
\uppercase{I}nstituci\'o \uppercase{C}atalana de \uppercase{R}ecerca  i 
\uppercase{E}studis  \uppercase{A}van{\c c}ats}}

\address{ICFO-Institut de Ci\`encies Fot\`oniques, 
E-08034 Barcelona, Spain\\
Institut f\"ur Theoretische Physik, Universit\" at  Hannover, D-30167 Hannover, Germany.\\
E-mail: Aditi.Sen@icfo.es\\
E-mail: Ujjwal.Sen@icfo.es\\
E-mail: Maciej.Lewenstein@icfo.es}

\author{Veronica AHUFINGER}
\address{Grup d'\'Optica. Universitat Aut\`onoma de Barcelona, E-08193 Bellaterra, Spain.\\
Institut f\"ur Theoretische Physik, Universit\" at  Hannover, D-30167 Hannover, Germany.\\
E-mail: Veronica.Ahufinger@uab.es}

\author{Marisa PONS}
\address{
Depart. de F\i sica Aplicada I.
Universidad del Pa\i s Vasco E-20600 Eibar, Spain.\\
E-mail: Marisa.Pons@ehu.es}

\author{Anna SANPERA$^{\dagger}$}
\address{Grup de F\i sica Te\`orica. Universitat Aut\`onoma de Barcelona, E-08193 Bellaterra, Spain.\\
Institut f\"ur Theoretische Physik, Universit\" at  Hannover, D-30167 Hannover, Germany.\\
E-mail: sanpera@ifae.es}

\maketitle

\abstracts{
We report our research on disordered complex systems using cold gases and
trapped ions, and address the possibility of using complex systems for quantum 
information processing. Two simple paradigmatic models of disordered complex
systems are revisited here. The first one corresponds to a  
short range disordered Ising Hamiltonian (spin glasses), 
which can be implemented with a Bose-Fermi (Bose-Bose) mixture 
in a disordered optical lattice.
The second model we address here is a long range disordered Hamiltonian, 
characteristic of neural networks (Hopfield model),
which can be implemented in a chain of trapped 
ions with appropriately designed interactions.}

\section{Introduction}
Complex many body systems are often characterized by 
structurally simple interactions, but complexity arises because 
the different terms or constraints appearing in the Hamiltonian 
compete one with another. 
If the system presents disorder, the Hamiltonian
is no longer translational invariant and depends 
locally on random parameters. When the system is not able 
to accommodate to all the constraints present in the Hamiltonian, 
it exhibits frustration. This leads
to the appearance of exotic phenomena, e.g. fractal, 
hierarchic, or ultrametric structures, distinct quantum phase 
transitions, etc\cite{sachdev}.
Over the last 40 years, disordered and 
frustrated systems have played a central role in  condensed matter 
physics and have posed some of the most challenging 
open questions of many body systems. 
Quenched disorder (i.e., frozen disorder) 
determines the physics of various phenomena, 
from transport and conductivity through localization, 
to percolation, spin glasses, neural networks, high Tc-superconductivity, etc.
The description of such systems is, however, extremely 
difficult, because it normally requires the averaging 
over each particular realization of the disorder. 
Systems which are not disordered but frustrated, 
lead very often to similar difficulties because, at low temperatures,  they are often 
characterized by an enormously large number 
of low energy excitations.

Recently, it has been shown that one can introduce local 
disorder and/or frustration in ultracold quantum gases 
in a {\bf{controlled}} way, using 
various experimentally feasible methods (for details see e.g.\cite{ahufinger}, and references therein), ranging from 
using several incomensurable optical lattices to trap the atoms,
or superimposing a speckle pattern in a regular optical lattice, 
or taking advantage of Feshbach resonances in 
fluctuating or inhomogeneous magnetic fields in order to induce 
a novel type of disorder that corresponds to random, 
or at least inhomogeneous nonlinear interaction couplings. 
Thus, different disorder and/or frustrated systems can be
conveniently prepared to study e.g. 
Bose glass\cite{damski}, Anderson localization\cite{aspect,arlt},
fermionic spin glasses\cite{anna} or quantum percolation\cite{anna},
kagom\'e lattices\cite{kagome} among others.
We have also recently investigated the possibilities offered by trapped ions
with engineered interactions\cite{wunderlich,porras} 
to model neural networks\cite{marisa}. A review of the 
different phases displayed by ultracold atomic gases 
in disordered optical lattices can be found in\cite{ahufinger}.

In this contribution, we present our approach to the study
of both, short and long range, disordered systems. 
In the former case, we focus on a spin glasses model\cite{ea}, 
i.e. short-range disordered magnetic systems which can
be simulated by Bose-Fermi mixtures in random potentials. In the case
of long range interactions, we study a neural 
network model simulated by a chain of trapped ions with appropriately 
designed interactions. 
In both cases, we examine the possibilities offered by those systems 
for quantum information tasks. In spite of the fact that using 
disordered systems to perform quantum
information processing seems to be an impossible task, 
at least two possible advantages arise immediately. First, 
these systems have typically a large number of different metastable (free) 
energy minima, as it happens in spin glasses (SG)\cite{parisi}. 
Such states might be used to store information distributed over the whole
system, similarly as in  neural network (NN) models\cite{amit}. 
The information is thus naturally stored in a redundant way, 
like in error correcting schemes\cite{birat-kando}. Second, 
in disordered systems with long range interactions, 
the stored information is robust: metastable states have quite large
basins of attraction in the thermodynamical sense. 
We have shown\cite{sensen} that in both models, 
short and long range, it is possible to generate
entanglement that survives over long times. 
Moreover, in the neural network
model, it is possible to store patterns that can be used as distributed
qubits over the whole system. Since the patterns 
are robust and act as attractor points in the energy diagram, these 
qubits can be partially destroyed by noise or any other non-desired
effect. The free evolution of the systems, however, retrieves the patterns back
and thus makes the qubits very robust.

\section{Short range disordered systems: Spin glasses}\label{sec:sping}

Spin glasses are random disordered systems with competing ferromagnetic and
antiferromagnetic interactions, which in dimensions $d>1$ present
frustration, since it is not possible to simultaneously accommodate 
all pairs of spins connected by a ferromagnetic (antiferromagnetic) bond. 
In the early 70's, Edwards and Anderson realized that the
essential physics of a spin glass does not lay 
in the details of their microscopic interactions, 
but rather in the competition between quenched ferro- and 
antiferro-magnetic interactions. To study the nature of spin glasses, they proposed  
a very simple short range disordered Ising Hamiltonian,
nowadays known as the Edwards-Anderson (E-A) model of spin glasses\cite{ea}: 
\begin{equation}
H_{E-A} = -\sum_{\left\langle ij\right\rangle }
J_{ij} \sigma_i^{z}\sigma_j^{z} -h^{z} \sum_i \sigma_i^z.
\label{spinglass}
\end{equation}
Here $\sigma_k^z$ denotes an Ising spin ($\pm 1$) 
at the \(k\)-th site, the $J_{ij}$'s describe nearest neighbor couplings for
an arbitrary lattice and $h^{z}$ 
is a magnetic field along the z-direction.
In the E-A model, the $J_{ij}$ couplings 
are given by independent random variables, which have  Gaussian probability
distributions with mean $\bar J=0$ 
and variance \(\Delta^2\). 
Since interactions are short range, a mean field theory cannot be 
used\cite{sherrington} 
and, traditionally, one has to rely on replica tricks\cite{parisi}, 
to do the appropriate average over the quenched disorder, in order to 
obtain the free energy $F$ of the system, and  derive 
the thermodynamical properties of the system from $F$.
A formally identical Hamiltonian as the one of Eq. (\ref{spinglass}) can be
derived from the Bose-Fermi (Bose-Bose) Hubbard Hamiltonian\cite{sachdev} 
(BFH, BH) describing a 
Bose-Fermi (Bose-Bose) mixture in an optical lattice with random disorder:
\begin{eqnarray}
&&H_{\mathrm{BFH}}=-\sum_{\left\langle ij\right\rangle }
(T_B b_{i}^{\dagger }b_{j}+T_F f_{i}^{\dagger }f_{j}+ {\rm h.c.}) \label{Hamiltonian} \\
&&+\sum_{i}\left[
\frac{1}{2}Vn_{i}(n_{i}-1)+Un_{i}m_{i}-\mu^B_i n_{i}-\mu^F_i m_i\right].
\nonumber
\end{eqnarray}
Here $b_{i}^{\dagger }$, $b_{j}$, $f_{i}^{\dagger }$, 
$f_{j}$ are the bosonic and fermionic
creation-annihilation operators,  $n_{i}=b_{i}^{\dagger
}b_{i}$, $m_{i}=f_{i}^{\dagger }f_{i}$ are the number operators, 
and $\mu^B_i$ and $\mu^F_i$ are  the bosonic
and fermionic local chemical potentials, respectively. 
The BFH model describes: i) nearest neighbor (n.n.) 
boson (fermion) hopping, with an  
associated negative energy $-T_{B}$ ($-T_{F}$); 
ii) on-site repulsive boson-boson interactions with
an energy $V$; iii) on-site boson-fermion interactions with an
energy $U$, which is positive (negative) for repulsive
(attractive) interactions, and finally, iv) interactions with the external
inhomogeneous potential, with energies $\mu^B_i$ and $\mu^F_i$.
In the limit of equal tunneling for bosons and fermions ($T_{B}= T_{F}=T$) 
and a strong coupling regime ($T\ll U,V$), using a quasi-degenerate
perturbation theory up to the second order, an effective
Hamiltonian can be derived, which describes the dynamics of the Bose-Fermi mixture
in terms of composite fermions\cite{lewen,anna} 
made of one fermion  plus -s bosons  or one fermion plus s bosonic holes.
The annihilation operators of the 
composite fermions are given by\cite{lewen}:
\begin{eqnarray}
F_i & = & \sqrt{\frac{(\tilde n -s)!}{\tilde n !}} \left( b_i^\dagger \right)^s f_i \quad\mbox{for $s$ bosonic holes}\label{eq:composites1} \\
F_i & = & \sqrt{\frac{\tilde n!}{(\tilde n -s)!}} \left( b_i \right)^{-s} f_i \quad\mbox{for $-s$ bosons}.
\label{eq:composites2}
\end{eqnarray}
Using the above notation, the effective Hamiltonian reads:
\begin{equation}
H_\textrm{eff}=-\sum_{\langle i,j \rangle}t_{ij}(F^{\dagger}_iF_j+ h.c.)+
\sum_{\langle i,j \rangle}K_{ij}M_iM_j - \sum_i \bar{\mu_i} M_i
\label{Heff}
\end{equation}
where $M_i=F_i^{\dagger}F_i$.
In the limit of negligible tunneling between composites ($t_{ij}\approx 0$), 
the Hamiltonian reduces to the E-A Hamiltonian (Eq.(\ref{spinglass})) with
an effective inhomegenous magentic field given by $\bar{\mu_i}$.

Let us now address the generation and evolution
of nearest neighbors (n.n.) entanglement in this model. 
To deal with quantum information processing in the E-A model, we consider now
a quantum Ising model. Therefore, in the following 
$\sigma_k^z$ denotes the Pauli z-operator. 
In the short range Ising model without disorder,
it is possible to create cluster and graph states (i.e. entanglement) 
starting from an appropriate initial product state\cite{briegelnew}.  
Here we show that, while the disorder averaged density matrix 
of two neighboring spins remains always separable,
the disorder averaged entanglement (quantified by the logarithmic negativity\cite{VidalWerner})
converges with time to a finite value\cite{sensen}. 
The generation of entanglement as well as 
its evolution for arbitrary times in an Ising model without disorder
but with long-range interactions, 
has also been addressed in Ref.\cite{briegelnew}. 
Starting from a pure product state of the form
$|\Psi\rangle=\prod_{i}|+\rangle_i$, 
where \(\left| + \right\rangle = (\left|0\right\rangle + \left|1\right\rangle)/\sqrt{2}\),
we evaluate first the density matrix of the system 
after a finite time:
  \(\rho(t,\{J_{ij}\}) = \exp\{-iH_{E-A}t\}|\Psi\rangle\langle \Psi|\exp\{+iH_{E-A}t\}\). To calculate n.n. entanglement with respect the 
pair (i,j) we calculate first the reduced density matrix tracing over all 
spins except (i,j), and then we use logarithmic negativity 
as a measure of entanglement of the remaining two-qubit system. 
Finally we average over the disorder. 

Our results show that  (i) after a finite time, entanglement
converges to a finite amount (see Fig. 1) independently of the
mean $\bar J$ of the Gaussian distribution, although the short-time
dependece does depend on the mean, and  (ii) n.n. entanglement decays
exponentially with the number of neighbors of a given site, which in turn depends on
the configuration of the lattice we consider. For example, for a  1D chain, any pair of
neighboring lattice sites has 2 neighbors, a 2D honey comb lattice has 4, 
a 2D square lattice has 6, and a 3D square lattice has 10. 

\begin{figure}[ht]
\begin{center}
\includegraphics[width=0.7\linewidth]{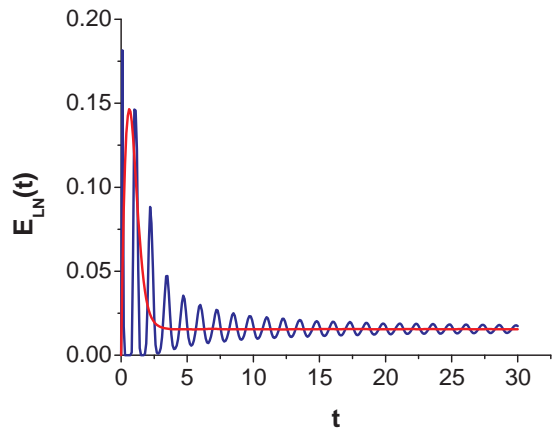}
\caption{Temporal behavior of n.n. averaged entanglement in a 
2D square lattice spin glass model. 
For a model with
$\bar J=0$, the logarithmic negativity $E_{LN}(t)$ quickly converges  to a constant value. 
For the case $\bar J=5$, $E_{LN}(t)$
exhibits damped oscillations, but again converging to 
the same value $\approx 0.0154$ as reached by the frustrated case. 
It is interesting to note that the dynamical behavior of the entanglement
depends on $\bar J$, although at large times, it converges to a fixed
value, independent of $\bar J$.}
\end{center}
\end{figure}
\section{Long range disordered systems: Neural Networks}\label{neuralnet}
Neural networks are paradigmatic models of parallel
distributed information processing\cite{amit,parisi}, 
and have been intensively studied by 
physicists since the famous paper by Hopfield\cite{hopfield}.
Following the models of Hopfield\cite{hopfield} and Little\cite{little}, 
a neuron can be viewed as an Ising spin with two possible states: an ``up" position $(S=+1)$ and a ``down" position $(S=-1)$, 
depending on whether the neuron has or has not fired an electromagnetic signal, in a given interval of time. 
The state of the network of $N$ neurons at a certain time is defined by the instantaneous configuration of 
all the spin variables $\lbrace S_i\rbrace$ at this time. The dynamic evolution of these states is 
determined by the interactions, $J_{ij}$, among neurons. The interaction are 
symmetric, so that
for any pair of neurons, $J_{ij}=J_{ji}$. Moreover, full connectivity is assumed, that is, 
every neuron can receive an input from any other one, and can send an output to it. 
The Hamiltonian of the system can be written as
\begin{equation}
H_{NN}=-\frac{1}{2}\sum_{i,j}J_{ij}S_iS_j.
\label{neural}
\end{equation}  
The interactions, $J_{ij}$, among neurons  are calculated {\em{a posteriori}}  by first fixing the configurations or 
patterns of spins to be stored in the network. 
In this way, the learning process is Hebbian, meaning that 
learning adjusts the network's weights 
such that its output reflects its familiarity with an input. 
The more probable an input, the larger the output will become (on average). 
Physically that means that these patterns will be learned if the system is able to accommodate 
them as attractors, so that they are stable in front of any single-spin flips and present a significant 
basin of attraction. 
Therefore, the interactions, $J_{ij}$, are defined in such a way 
that the local minima of the Hamiltonian are correlated 
with these configurations:
\begin{equation}
J_{ij}=\frac{1}{N}\sum _{\mu=1}^p \xi_i^{\mu}\xi_j^{\mu}.  
\label{int_neural}
\end{equation} 
Here $i\not= j$. The $p$ sets of \(\lbrace \xi_{i}^{\mu}\rbrace\) (where 
each \(\xi_i^\mu\) can be \(\pm 1\)) 
are the patterns that we wish to be fixed by the learning process. 
Despite the fact that the interactions have been constructed to guarantee that certain specified patterns are fixed 
points of the dynamics, the non-linearity of the dynamical process induces additional attractors, 
the so-called spurious states.

Recently it has been demonstrated that a linear chain of harmonically 
trapped ions can be appropiately designed, by applying either an external 
magnetic field\cite{wunderlich} or external lasers\cite{porras}, to describe
a spin sytem with long range interactions:
\begin{equation}
H=-\frac{1}{2}\sum_{ij}J_{ij}\sigma_i\sigma_j+\sum_{i}B'\sigma_i,
\label{neural2}
\end{equation}
where
\begin{equation}
J_{ij}=\frac{F^2}{m}\sum_n \frac{M_{i,n}M^{j,n}}{\omega^2_{n}}.
\label{int}
\end{equation}
The $M$'s are the unitary matrices that diagonalize 
the vibrational Hamiltonian. With these assumptions, 
Eq. (\ref{neural}) and  Eq. (\ref{neural2}) 
have the same form, and the possibility of implementing a classical neural 
network using a linear chain of ions arises. 
Also, comparison between Eq. (\ref{int_neural}) and Eq. (\ref{int}), 
indicates that the network configuration in the ions trap case 
will be given by the vibrational modes of the chain ($M_{i,n}$) and their
eigenvalues ($\omega_n$). 
The vibrational modes are determined through the harmonic displacements 
of the ions around their equilibrium positions, when the trapping potential 
is balanced by the Coulomb interactions between the ions. Thus, in our 
model, the sign of the displacement of each ion with respect to 
its equilibrium position, is associated with an Ising spin.
To reproduce the NN model (Eq. (\ref{int_neural})), and thus to 
be able to store different patterns, the vibrational modes 
should be, ideally, almost degenerate in energy
and possess large basins of attraction. In other words, the patterns
should correspond  
to sufficiently different configurations of the spins, so that each 
configuration is stable in front of random spin flips of several 
of its components. 
To achieve the above situation, either the vibrational spectrum is modified 
by changing the shape of the trapping potential, or an external 
longitudinal magnetic field (along the axial frequency) is used.
We obtain the best possible results 
concerning the number of stored patterns, if we use for the 
ions' trap, a confining potential of the form $V(x)=A|x|^{0.5}$,
 without any additional external magnetic field. 
In this case, we can store up to 4 patterns (2 patterns plus their reverse ones) in a 20 ions chain 
(for details see\cite{marisa}). Notice that due to the 
fact that the interactions are now given by the vibrational modes, our system
does not correspond exactly to the Hopfield model of neural networks, and
therefore, the system is not able to learn the same number of patterns
as the Hopfield one.

Let us now move to the entanglement properties of a quantun neural network. As
before, we replace the classical Ising spins, $\{S_i\}=\pm 1$, by Pauli
operators $\sigma_i^z$.  We apply here the same procedure as in 
Section 2  and consider as initial state 
a pure product state of the form 
$|\Psi\rangle=\prod_{i}|+\rangle_i$.
The entanglement of any two spins 
is calculated by evaluating first the time evolved density matrix  
\(\rho(t,\{J_{ij}\}) = \exp\{-iH_{NN}t\}|\Psi\rangle\langle \Psi|\exp\{+iH_{NN}t\}\), and then tracing over all
subsystems except ${i,j}$, and finally performing the proper average 
over the disorder (for details see\cite{sensen}). 
We have shown that 
(i) there is an efficient way to calculate bipartite as well as multipartite states in this model,
and (ii) entanglement displays 
collapses and revivals as a function of time and number of ions.

\end{document}